\documentclass[11 pt]{article}
\usepackage{geometry} 
\usepackage{graphicx}
\usepackage{amssymb}
\usepackage{epstopdf}
\usepackage{amsmath}
\usepackage{pdflscape}
\usepackage{afterpage}
\usepackage{times}
\usepackage{caption}
\usepackage{indentfirst}
\usepackage{float}
\usepackage{xcolor}
\usepackage{bm}
\usepackage[cm]{fullpage}
\usepackage[labelfont=bf]{caption}
\usepackage{nopageno}

\title{Photothermal optomechanics in superfluid helium coupled to a fiber-based cavity.}
\author
{A.D. Kashkanova,$^{1}$ A.B. Shkarin,$^{1}$ C. D. Brown,$^{1}$ N. E. Flowers-Jacobs,$^{1}$ L. Childress,$^{1,3}$\\
S.W. Hoch,$^{1}$ L. Hohmann,$^{3}$ K. Ott,$^{3}$ J. Reichel,$^{3}$ and J. G. E. Harris$^{1,4}$\\
\\
\normalsize{$^{1}$Department of Physics, Yale University, New Haven, CT, 06511, USA}\\
\normalsize{$^{2}$Department of Physics, McGill University, 3600 Rue University, Montreal, Quebec H3A 2T8, Canada}\\
\normalsize{$^{3}$Laboratoire Kastler Brossel, ENS/UPMC-Paris 6/CNRS, F-75005 Paris, France}\\
\normalsize{$^{4}$Department of Applied Physics, Yale University, New Haven, CT, 06511, USA}\\
}
\date{} 
\begin{document}
\maketitle
\section{Abstract}
Presented in this paper are measurements of an optomechanical device in which various acoustic modes of a sample of superfluid helium couple to a fiber-based optical cavity. In contrast with recent work on the paraxial acoustic mode confined by the cavity mirrors \cite{Kashkanova2016}, we focus specifically on the acoustic modes associated with the helium surrounding the cavity. This paper provides a framework for understanding how the acoustic modes depend on device geometry. The acoustic modes are observed using the technique of optomechanically induced transparency/amplification. The optomechanical coupling to these modes is found to be predominantly photothermal.

\section{Introduction}
Incorporating fluid into an optomechanical system can be beneficial in a number of ways. First, it can provide new avenues for studying the fluid's properties \cite{Bahl2013, HyunKim2013a}. Second, a fluid with high thermal conductivity can be used to thermalize a mechanical element, allowing higher optical powers to be used \cite{Sun2013}. Third, a fluid can be used as a mechanical element, which simplifies the assembly process, as the fluid can conformally fill or coat an electromagnetic resonator \cite{Colgate1991,Harris2015}. 

There are two distinct approaches to using fluid in an optomechanical system: immersing a solid resonator into fluid \cite{Harris2015, Gil-Santos2015}, or filling a hollow resonator with fluid \cite{Kashkanova2016,Bahl2013,HyunKim2013a, Lorenzo2014}. If one wishes to perform experiments at cryogenic temperature (which is necessary to achieve quantum behavior in almost all optomechanical systems), liquid helium is the only choice of fluid, since it does not solidify under its own pressure. In addition it has a variety of useful properties for optomechanical applications, such as very low optical absorption \cite{Seidel2002189}, high thermal conductivity \cite{Greywall1981}, and acoustic loss proportional to $T^4$ \cite{Abraham1969}, which becomes low at dilution refrigerator temperatures.

In this paper we present investigations of the coupling between superfluid helium and a fiber cavity. In contrast with our previous work \cite{Kashkanova2016}, we do not focus on the paraxial acoustic mode confined by the cavity mirrors and co-located with the optical modes, but rather study lower frequency modes defined by the fiber cavity and the surrounding components. We find that the coupling to those modes is not mediated through radiation pressure or electrostriction, but is predominantly photothermal. Photothermal coupling is widely studied, and arises from heating due to optical absorption \cite{Aspelmeyer2014,Vogel2003,Zalalutdinov2001,Metzger2004}. In the system described below, we have observed the change in the spring constant and the linewidth of the acoustic modes, using the technique of optomechanically induced transparency/amplification.

\section{Methods}
\begin{figure}[h!] 
\centering
\includegraphics[width=0.9\textwidth]{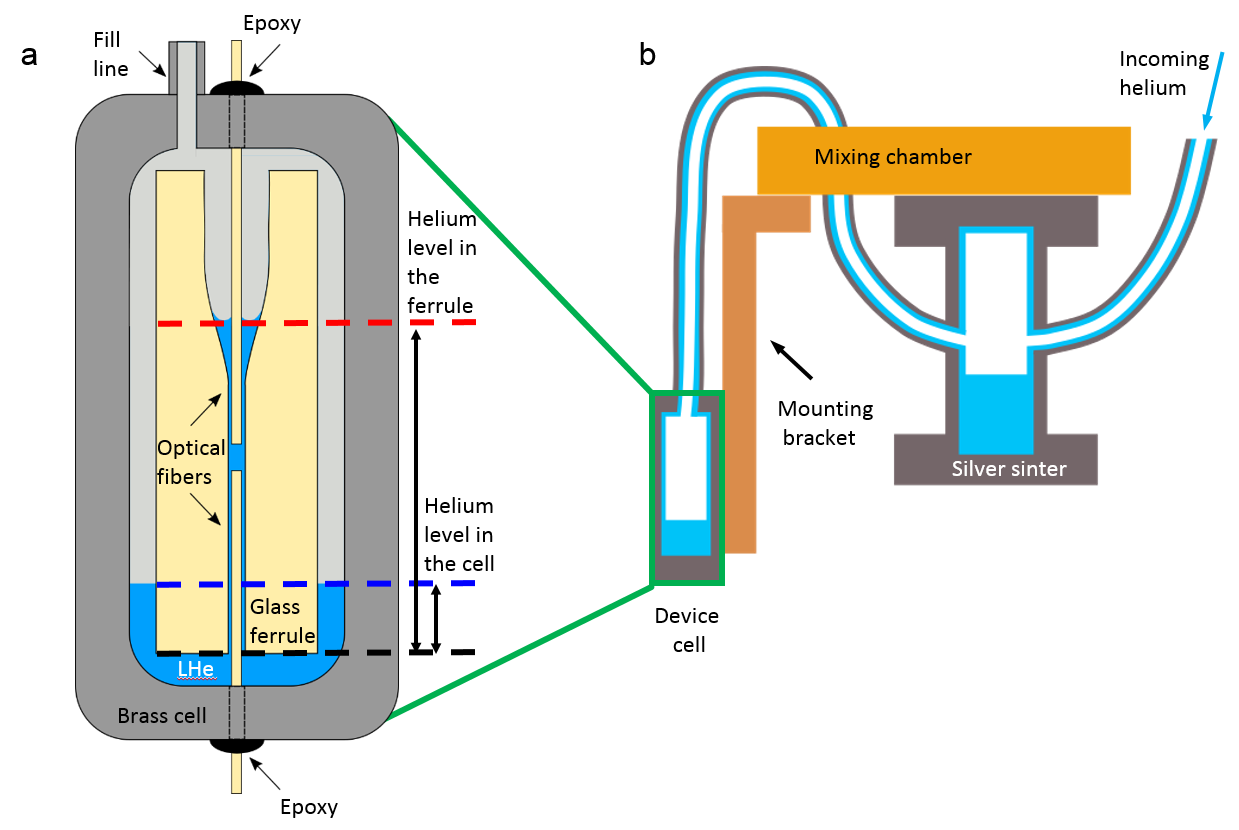} 
\caption{\textbf{The experimental setup}. 
\textbf{a)} A schematic drawing of the device. The optical cavity is formed between two glass fibers (yellow), aligned in a glass ferrule (yellow). The ferrule is epoxied into a brass cell (gray). The optical fibers are epoxied to the brass cell as well, forming superfluid helium tight seals. The helium (blue) is delivered via the fill line shown at the top of the cell. The helium level in the cell is measured from the bottom of the ferrule (dashed black line) and is shown with a dashed blue line. The helium level in the ferrule is higher due to capillary action and is shown with a dashed red line. 
\textbf{b)} A schematic drawing of the helium delivery system at the mixing chamber (MC) of the dilution refrigerator. The helium (blue) enters through a capillary (shown on the right). It first condenses in a silver sinter (gray), and then flows via Rollin film into the cell. The cell is attached to a gold-plated OFHC copper mounting bracket, which is attached to the MC.}
\label{fig:1}
 \end{figure}

The device, shown schematically in the Figure~\ref{fig:1}(a), is located in a cell made of brass. A glass ferrule with $133\pm 5$ $\mu$m diameter bore is epoxied inside the cell. The bore is funneled on one end. An optical cavity is created by inserting a pair of 125 $\mu$m diameter fibers into the bore \cite{Flowers-Jacobs2012}. Each fiber face (the cleaved end of the fiber) has an indentation created by CO$\mathrm{_2}$ laser ablation \cite{Hunger2012,Hunger2010a}. The depth of the indentation is $\simeq1.5$ $\mu$m and the radii of curvature are: $R_1=282$ $\mu$m and $R_2=409$ $\mu$m. On each fiber end, a Distributed Bragg Reflector (DBR) optical coating has been deposited \cite{Rempe:92}. The optical transmission is 103 ppm for the input mirror and 10 ppm for the back mirror, which results in a single sided cavity. The mirror separation is $L=84$ $\mu$m and the optical cavity linewidth is $\kappa=2\pi\times54$ MHz. The input coupling is $\kappa_{\mathrm{in}}=2\pi\times27$ MHz. 

The mounting of the cell is shown in Figure~\ref{fig:1}(b). The cell is mounted to the mixing chamber (MC) of a dilution refrigerator, which is kept at temperature $T<100$ mK. The cell is filled through a $1.5$ mm outer diameter (500 $\mu$m inner diameter) stainless steel capillary. The capillary is wound around copper bobbins at each stage of the refrigerator for thermalization. At the MC, there is a sintered silver heat exchanger which is used to thermalize the incoming helium to the MC temperature.

The device is filled by adding small doses of helium. The time when the dose is added and the amount of helium in each dose are recorded. Using this information, the level of helium in the cell can be calculated, as described in section \ref{s:filling}.

The resonant frequency of the optical cavity is monitored in order to determine when the cavity is filled. Filling the cavity changes its effective length by a factor of $1.028$ (the index of refraction of LHe) \cite{Donnelly1998}, changing the frequency of the cavity modes correspondingly. Once the cavity is filled, the OptoMechanically Induced Transparency (OMIT) \cite{Weis2010} is used to observe the acoustic modes. Using this technique, two distinct families of acoustic modes are observed: low frequency modes ($20-300$ KHz) and high frequency modes ($2-20$ MHz). Some of the low frequency modes show strong dependence on the helium level, from which we conclude that the mode profiles extend beyond the cavity volume into the helium sheath between the fiber and the ferrule, and into the funnel. Those modes are referred to as ``ferrule modes''. In contrast, the high frequency modes are independent of the helium level and have frequencies consistent with the radial acoustic modes of a cylinder of helium. Those modes are referred to as ``radial modes''. 


\subsection{Filling the device}
\label{s:filling}
As described above, helium is added to the cell in discrete doses. Each dose at room temperature contains $\simeq13.4$ cm$^3$ helium gas at a pressure of $\simeq1100$ mbar. The doses are added at a rate of approximately one every three minutes. The doses are first released into a liquid nitrogen cold trap, then, after one minute, are added to the cell. Due to the geometry of the system (shown in Figure \ref{fig:1}(b)), helium first condenses inside the volume containing the sintered silver heat exchanger (which is a local gravitational minimum), and then ``creeps'' as a Rollin film \cite{Rollin1939219} along the walls of a connecting capillary to fill the cell.
The film thickness $d$ is given by \cite{Pobell1992}
\begin{equation}
d=\eta h^{-1/3}
\end{equation}
Here $\eta= 6.5\times10^{-9}$ m$^{4/3}$ and $h$ is the height above the helium level.
Before helium starts to condense in the cell, all parts of the system (the sintered silver heat exchanger, the cell, the capillary connecting the sintered silver heat exchanger to the cell) need to be covered by the superfluid Rollin film. The surface area of the sintered silver heat exchanger is specified to be $\simeq10$ m$^2$ \cite{FRANCO1984477} which is at least two orders of magnitude larger than combined surface area of other parts of the system. The volume of liquid helium necessary to cover the sintered silver heat exchanger with Rollin film is estimated to be $0.16$ cm$^3<V_{\mathrm{film}}<0.38$ cm$^3$. Both upper and lower bounds are found by assuming constant film thickness throughout the sinter. The lower bound results from assuming the bulk helium level to be located in the device (6 cm below the top of the sinter), while the upper bound results from assuming the bulk helium level to be located in the volume containing the sinter (0.5 cm below the bottom of the sinter).

As soon as the amount of helium in the system exceeds $V_{\mathrm{film}}$, helium starts condensing at the bottom of the volume containing the sintered silver heat exchanger and starts flowing into the cell at a rate $\dot{V}$ \cite{Pobell1992}:
\begin{equation}
\dot{V}=2\pi R_\mathrm{c}d_{\mathrm{thin}} v_{\mathrm{crit}} 
\label{eq:dotv}
\end{equation}
Here, $R_\mathrm{c}$ is the inner radius of the capillary, $d_{\mathrm{thin}}$ is the film thickness at its thinnest point, and $v_{\mathrm{crit}} \simeq 30$ cm/s is the critical superfluid velocity \cite{Pobell1992}. The highest point of the capillary is located $\simeq20$ cm above the minimum liquid level in the sinter, resulting in the minimum film thickness $\simeq14$ nm. Therefore the rate of filling the device is $\dot{V}=4\times10^{-4}$ cm$^3$/min. If, at any point, there is no helium accumulated at the bottom of the sintered silver heat exchanger volume, helium stops flowing into the device.

Using this model, we can describe the volume of liquid helium accumulated in the device as a function of time. Using a CAD model of the cell, the helium level in the cell can be modeled as a function of the volume of helium accumulated in the device. This is shown with a blue line on Figure \ref{fig:2}(a).

\subsection{Capillary action}
\label{s:capact}
Since the cavity is located inside the ferrule, the helium level in the ferrule is of interest. The helium level in the ferrule is higher than the helium level in the cell due to capillary action \cite{Wang2012}. In what follows, we calculate the helium level in the ferrule as a function of the helium level in the cell.
\subsubsection{Hollow tube of constant radius}
To understand capillary action in the ferrule, consider first a simple model: a thin tube of radius $r$ submerged in a fluid bath. Assume that the pressure above the bath is zero. The surface tension is $\sigma$ and the contact angle is $\theta_c$, as shown in the inset of the figure \ref{fig:2} (b). The height $h^*$ to which the fluid rises can be determined by balancing gravitational potential energy and interfacial energy, which is done done by minimizing the free energy. The free energy of the system is given by:
\begin{equation}
F(h)=\int_V\rho g z dV-\int_A\sigma \cos\theta_c dA
\end{equation}
Here $V$ is the volume of the fluid in the capillary above the bath level, and $A$ is the area over which helium is in contact with glass above the bath level. The values $\rho$ and $g$ are the density of the fluid and the gravitational constant. 
In the case above, ignoring the meniscus, we arrive at the equation:
\begin{equation}
F(h)=\rho g \pi r^2\int_0^h z dz -\sigma\cos\theta_c 2 \pi r h=\rho g \pi r^2 \frac{h^2}{2} -\sigma\cos\theta_c 2 \pi r h
\end{equation}
The height of the fluid in the capillary is then:
\begin{equation}
h^*=\frac{2\sigma \cos\theta_c}{\rho g r}
\end{equation}
For a $133$ $\mu$m diameter capillary submerged in superfluid helium ($\rho=145$ kg/m$^3$ \cite{Donnelly1998},$\theta_c=0^{\circ}$ \cite{Eckardt1977} and $\sigma=3.78\times10^{-4}$ J/m$^2$ \cite{Eckardt1977}, the fluid rises by $h^*=8$ mm.

\subsubsection{Hollow axissymmetric tube of arbitrary shape}
The ferrule is a hollow axissymmetric tube of arbitrary shape: $r=r(z)$. Here $r$ is the distance from the axis of the ferrule and $z$ is the distance from the bottom of the ferrule. The free energy is:
\begin{equation}
F(h)=\rho g \pi\int_0^h r(z)^2z dz -\sigma\cos\theta_c 2 \pi\int_0^h r(z)\sqrt{1+\left(\left.\frac{dr}{dz}\right|_z\right)^2}dz
\end{equation}
In addition, there is a fiber in the center of the ferrule, which is modeled as a solid tube of constant radius $r_{\mathrm{fib}}$. The free energy then is:
\begin{equation}
F(h)=\rho g \pi\int_0^h \left [r(z)^2z -r_{\mathrm{fib}}^2h\right] dz -\sigma\cos\theta_c 2 \pi \int_0^h\left[ r(z)\sqrt{1+\left(\left.\frac{dr}{dz}\right|_z\right)^2}+r_{\mathrm{fib}}\right]dz
\end{equation}
The derivative of the free energy with respect to $h$, evaluated at $h^*$ is:
\begin{equation}
\left.\frac{dF}{dh}\right|_{h^*}=\rho g \pi\left [r(h^*)^2-r_{\mathrm{fib}}^2\right] h^* - \sigma \cos\theta_c 2 \pi \left[r(h^*)\sqrt{1+\left(\left.\frac{dr}{dz}\right|_{h^*}\right)^2}+r_{\mathrm{fib}}\right]=0
\label{eq:free energy der}
\end{equation}
Equation \ref{eq:free energy der} can be used to find the height of the capillary rise for the ferrule with a fiber, provided the ferrule profile is known. 

\begin{figure}[h!] 
\centering
\includegraphics[width=0.9\textwidth]{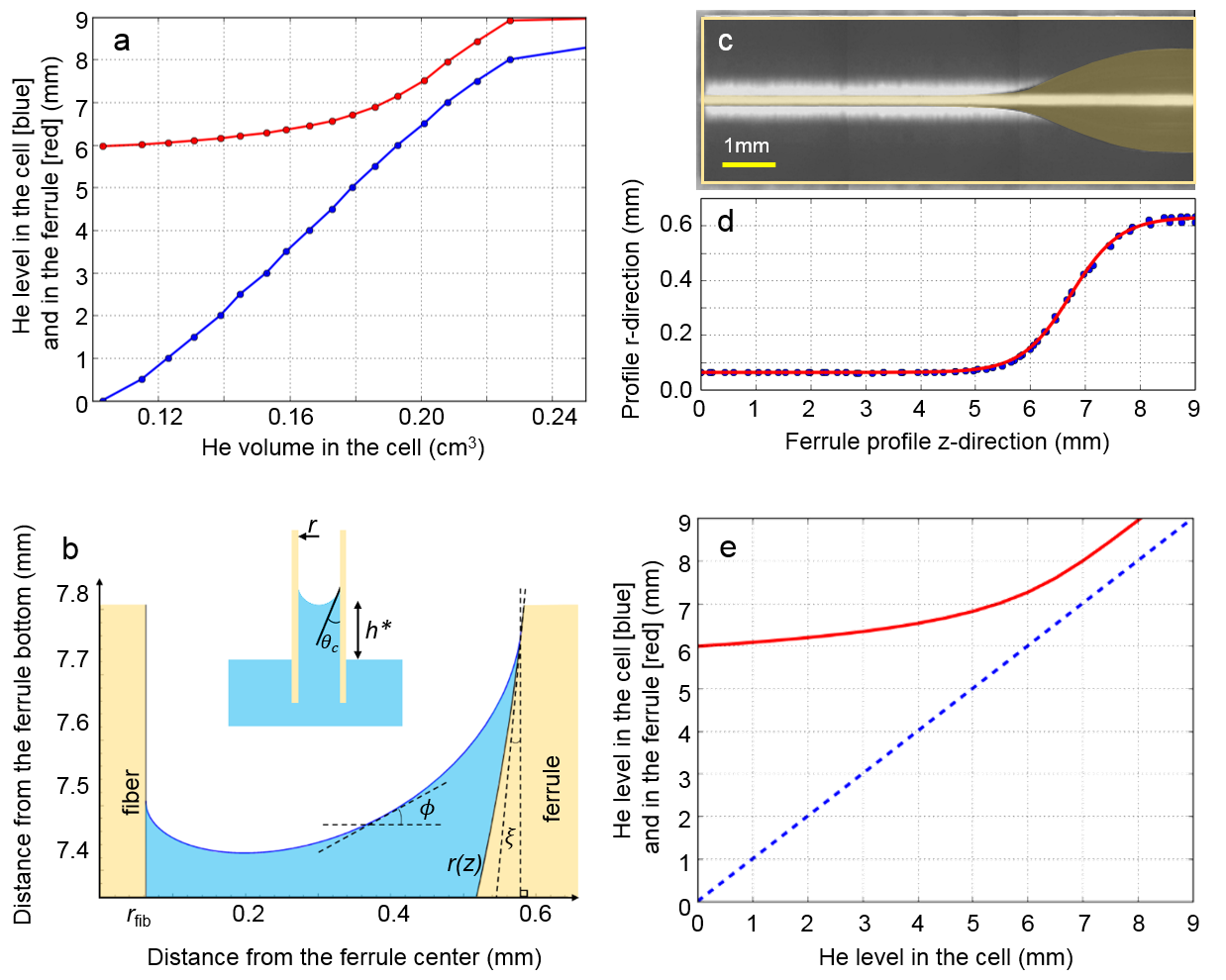} 
\caption{\textbf{Helium level in the ferrule}.
\textbf{a)} Helium level in the cell (blue) and in the ferrule (red) as a function of the volume of helium accumulated in the device (blue). Zero on the vertical axis corresponds to the bottom of the ferrule. Helium level in the cell for different volumes of helium accumulated in the cell was calculated using CAD model of the cell. Helium level in the ferrule was calculated using the blue points and the results from panel (e).
\textbf{b)} 
\textbf{Inset:} A schematic drawing of a cylindrical capillary (yellow) of radius $r$, submerged in liquid. The contact angle is $\theta$. Due to capillary action the liquid rises to height $h^*$. 
\textbf{Main figure:} A plot of the superfluid helium meniscus (blue) formed in the space between the fiber (yellow) and the ferrule (yellow), calculated using the procedure described in section \ref{s:capact}. The fiber has constant radius $r_{\mathrm{fib}}$. The ferrule profile is described by $r(z)$. The ferrule profile makes an angle $\xi(z)$ with the vertical. The angle $\phi$ is the slope angle of the liquid. 
\textbf{c)} A photograph of the ferrule. Yellow shading indicates the funnel and the bore.
\textbf{d)} The ferrule profile extracted from the photograph (c) is shown using blue dots. The red line is a fit of equation \ref{eq:tan} to the profile. 
\textbf{e)} The helium level in the ferrule plotted as a function of helium level in the cell (red solid line). The blue dashed line illustrates unity slope. As soon as helium touches the bottom of the ferrule, it gets sucked in by the capillary action to a height of 6 mm, which is where the bore starts to widen into the funnel.}
\label{fig:2}
 \end{figure}

A photo of the ferrule is shown in the Figure \ref{fig:2}(c). The ferrule is made of borosilicate glass with index of refraction $n=1.52$. The length and diameter of the ferrule are 9 mm and 3 mm respectively. This information can be used to correct the image and extract the profile of the ferrule bore. The extracted profile is shown in Figure \ref{fig:2}(d) with blue circles. In order to evaluate equation \ref{eq:free energy der} analytically, the profile is fit to the function: 
\begin{equation}
r(z)=A\tanh\left(\frac{x-B}{C}\right)+D
\label{eq:tan}
\end{equation}
The following values are found for the fit parameters: $A=284$ $\mu$m, $B=6739.5$ $\mu$m, $C=890.4$ $\mu$m, $D=347$ $\mu$m. The fit is shown in Figure \ref{fig:2}(d) with a red line.

In addition, helium forms a meniscus in the ferrule, which is shown in Figure \ref{fig:2}(b). Its shape is calculated following \cite{Pozrikidis2009a}. The shape of the interface is described in parametric form by the slope angle $\phi$:
\begin{equation}
z=Z(\phi)
\end{equation}
\begin{equation}
r=R(\phi)
\end{equation}
Two differential equations describe the shape of the meniscus:
\begin{equation}
\frac{dR(\phi)}{d\phi}=\frac{\cos\phi}{Q(\phi)}
\label{eq:dr}
\end{equation}
and
\begin{equation}
\frac{dZ(\phi)}{d\phi}=\frac{\sin\phi}{Q(\phi)}
\label{eq:dz}
\end{equation}
Where $Q(\phi)$ is given by 
\begin{equation}
Q(\phi)=\frac{Z(\phi)}{l^2}-\frac{\sin\phi}{R(\phi)}
\end{equation}
Here, $l$ is the capillary length: $l=\sqrt{\sigma/\rho g}$. For a constant-radius hollow capillary, equations \ref{eq:dr} and \ref{eq:dz} are solved subject to the following boundary conditions:
\begin{equation}
Z(0)=h_m^*
\label{eq:zini}
\end{equation}
\begin{equation}
R(0)=0
\label{eq:rini}
\end{equation}
Here, $h_m^*$ is the height of the lowest point of the meniscus. The height $h_m^*$ that satisfies the condition $R(\theta_c)=r$ is then found. 

For the system used in the experiment (a ferrule whose profile is described by a function $r(z)$ with a fiber with radius $r_{\mathrm{fib}}$ in the center), the initial conditions become: 
\begin{equation}
Z(-\pi/2-\theta_c)=h_m^*, 
\end{equation}
\begin{equation}
R(-\pi/2-\theta_c)=r_{\mathrm{fib}}
\end{equation}
$h_m^*$ is then found to satisfy the two conditions below:
\begin{equation}
R(\pi/2-\theta-\xi(z))=r(Z(\theta_c- \xi(z)))
\end{equation}
\begin{equation}
\arctan(r’(Z(\pi/2-\theta_c- \xi(z))))= \xi(z)
\end{equation}
Here, $\xi(z)$ is the angle of the ferrule profile, as shown in the Figure \ref{fig:2}(b). 
The solution to equation \ref{eq:free energy der} is used an the initial guess for $h_m^*$.

The helium level in the ferrule, calculated using the procedure outlined above, is shown in Figure \ref{fig:2}(e) with a red line. The blue dashed line is a line with unity slope. From Figure \ref{fig:2}(e) it is clear that as soon as superfluid helium reaches the bottom of the ferrule, the helium level inside of the ferrule goes up to $6$ mm, filling the optical cavity and making it possible to observe the acoustic modes. The relationship between the helium level in the cell and helium level in the ferrule allows us to plot the helium level in the ferrule as a function of liquid helium volume in the cell (shown in Figure \ref{fig:3}(a) with a red line).


\section{Mode shapes}
\subsection{Optical modes}
The intensity of the TEM$_{00}$ optical cavity mode employed in the experiment is approximately proportional to:
\begin{equation}
I(r,z)\propto\sin\left(\frac{2\pi z}{\lambda_{\mathrm{opt}}}\right)^2e^{-\frac{2r^2}{w^2}}
\label{eq:Igood}
\end{equation}
Here $\lambda_{\mathrm{opt}}=1,504$ nm is the optical wavelength in liquid helium and $w \simeq3.5$ $\mu$m is the mode field radius \cite{Lasers}. Since the indentations on the fiber faces are not always centered on the fibers, and the fibers are possibly not centered in the ferrule, the optical mode (confined by the indentations) can be slightly offset from the center of the radial acoustic mode (confined by the inner walls of the ferrule). Assuming the existence of such offset, $x_0$, the intensity profile can be modeled in the following manner:
\begin{equation}
I(x,y,z)\propto\sin\left(\frac{2\pi z}{\lambda_{\mathrm{opt}}}\right)^2e^{-\frac{2((x-x_0)^2+y^2)}{w^2}}
\label{eq:Ioffset}
\end{equation}
The offset of the indentation from the fiber centers was measured to be $ x_0 \simeq1-3$ $\mu$m.

\subsection{Radial acoustic modes}
\begin{figure}[h!] 
\centering
\includegraphics[width=0.6\textwidth]{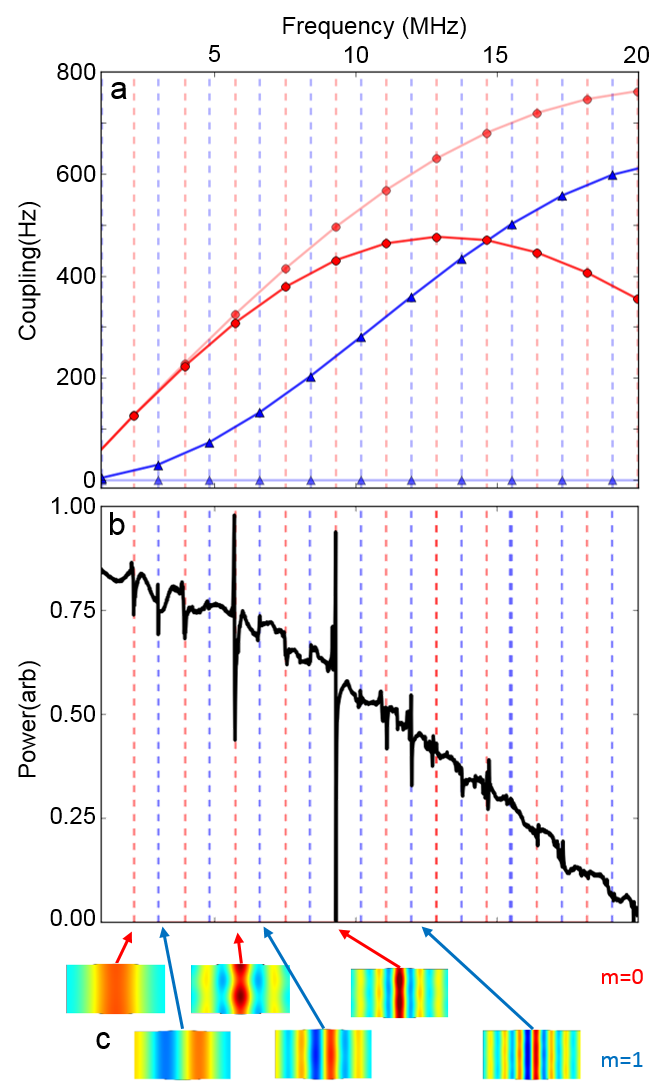} 
\caption{\textbf{The radial modes}. 
\textbf{a)} The electrostrictive coupling rate $g_0$ to the radial modes of different frequencies. The red dashed and blue dashed vertical lines show the frequencies of the radial modes with $m=0$ and $m=1$ respectively, calculated using equation \ref{eq:omegamn}. The faint red (blue) line connecting the red circles (blue triangles) shows the coupling to $m=0$ ($m=1$) radial modes, calculated using methods in section \ref{ss:Optomechanical coupling}, assuming the optical and acoustic modes are perfectly aligned. The bright red (blue) line connecting the red circles (blue triangles) shows the coupling to $m=0$ ($m=1$) radial modes, assuming the optical and acoustic modes are offset by $3$ $\mu$m. The misalignment leads to decrease in coupling to $m=0$ mode and to emergence of coupling to $m=1$ mode. 
\textbf{b)} The red dashed and blue dashed vertical lines show the frequencies of the radial modes with $m=0$ and $m=1$ respectively, found using COMSOL simulations. They match with the lines of panel (a), demonstrating that the deviations of the shape of the space between the fibers from the ideal cylinder do not influence the frequencies of the modes significantly. The OMIT/OMIA signal for the intracavity beatnote frequency in range $1-20$ MHz is shown with a thick black line. The overall downward slope is due to the cavity filtering. The measured mode frequencies match well with the predicted frequencies for $m=0$ and $m=1$ modes. 
\textbf{c)} The mode profiles for some of the modes, calculated using COMSOL. }
\label{fig:3}
 \end{figure}

The acoustic radial modes in the LHe can be modeled as solutions of the time independent wave equation in cylinder of radius $R$ with zero flux on all surfaces and with zero longitudinal number, that is independent of $z$.
\begin{equation}
p_{mn}(r,\theta)=J_m(\alpha_{mn} r/R) \cos m\theta
\end{equation}
Here $J_m$ is the $m^{\mathrm{th}}$ Bessel function of the first kind, $\alpha_{mn}$ is the $n^{\mathrm{th}}$ zero of the derivative of $J_m$. The frequencies of the modes are:
\begin{equation}
\omega_{mn}=\frac{v\alpha_{mn}}{ R}
\label{eq:omegamn}
\end{equation}
Here $v=238$ m/s is the speed of sound in helium and $R=67$ $\mu$m.
The frequencies of the modes are shown in the Figure \ref{fig:3}(a) with red dashed lines ($m=0$) and blue dashed lines ($m=1$).

To check if deviations of the geometry from an ideal cylinder have noticeable effect, the frequencies of the radial modes are calculated by using finite element modeling software (COMSOL) to solve the wave equation for a cylinder of helium with length $L=84$ $\mu$m and radius $R=67$ $\mu$m with the $1.5$ $\mu$m deep indentations for the mirrors with appropriate radii of curvature. The wave equation is solved for both $m=0$ and $m=1$ cases. The boundary condition is ``zero flux'' on all the boundaries. The frequencies obtained via COMSOL simulations are in good agreement with frequencies calculated for a perfect cylinder, as can be seen by comparing red ($m=0$) and blue ($m=1$) lines in the Figure \ref{fig:3}(b) with the corresponding lines found analytically and shown in Figure \ref{fig:3}(a). The profiles for some of the modes are shown in Figure \ref{fig:3}(c). 

\subsection{Ferrule acoustic modes}
The frequencies of the ferrule modes are found by implementing the whole geometry of the ferrule, described by equation \ref{eq:tan}, fibers and meniscus in COMSOL and solving the wave equation for this geometry. All of the helium-glass boundaries are set to be ``zero flux'' boundaries. The bottom end of the ferrule is open, so it requires zero-pressure boundary condition (``Dirichlet boundary condition'') and the top surface of the helium (helium-vacuum boundary), which has the meniscus shape as shown in Figure \ref{fig:2}(b), is described by a ``free surface'' boundary condition. 

\begin{figure}[h!] 
\centering
\includegraphics[width=0.8\textwidth]{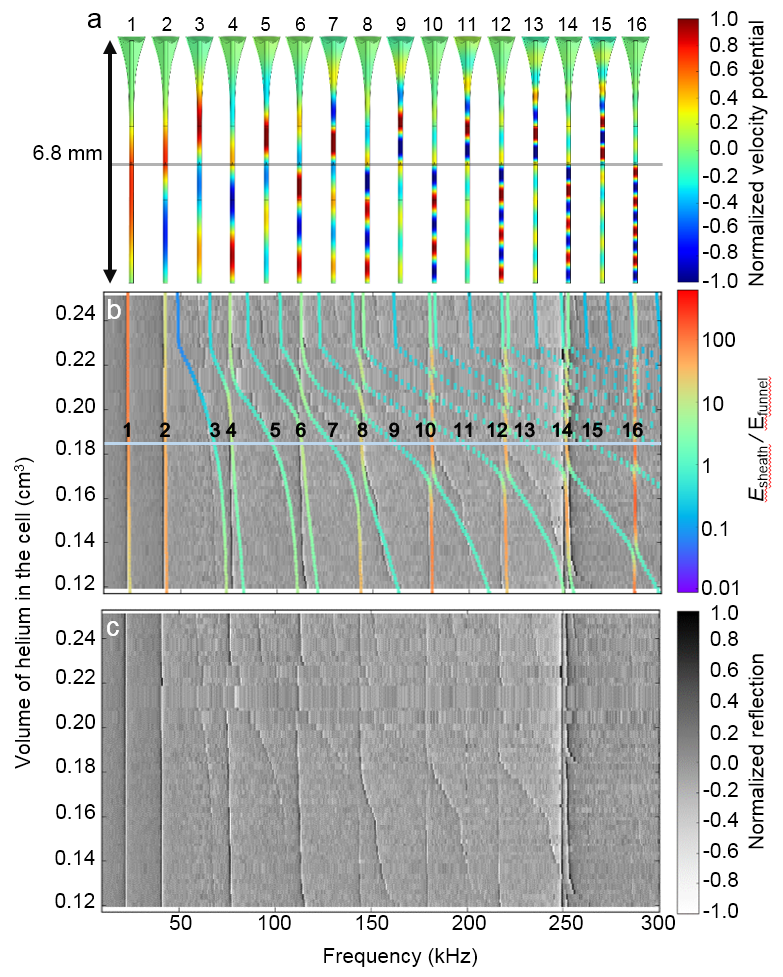} 
\caption{\textbf{The ferrule modes}. 
\textbf{a)} Several mode profiles of the helium in the ferrule when the helium level is 6.8 mm, calculated using COMSOL. The color indicates the normalized velocity potential. Some of the modes (e.g. 10,12,14,16) are mostly localized in the sheath of helium surrounding the fibers in the bore below the cavity. Some other modes (e.g. 9,12,13,15) are mostly localized in the helium above the cavity. The gray line shows the position of the cavity.
\textbf{b)} Solid lines: the frequencies of the modes as a function of the volume of helium accumulated in the device. The color indicates the ratio of energy stored in the sheath to the energy stored in the funnel. The modes that are mostly confined in the sheath do not change in frequency as the more helium accumulates in the cell. The modes that extend into the funnel show a decrease in frequency as the helium level rises. The light blue line is at 0.182 cm$^3$ (which corresponds to 6.8 mm of helium in the ferrule and hence the simulations is panel (a)). The black numbers are located at the intersections of the light blue line with the colored lines correspond to the numbers in panel (a). Density plot: the measurements also shown in (c).
\textbf{c)} Density Plot of the OMIT/OMIA signal for intracavity beatnote frequencies in range 10-300 kHz as a function of the volume of helium condensed in the device.}
\label{fig:4}
 \end{figure}

Since it was experimentally observed that the frequencies of some of the ferrule modes change with the level of helium, we simulate the mode frequencies vs. the helium level in the ferrule. Figure \ref{fig:4}(a) shows the profiles of the modes when the helium level in the ferrule is $6.8$ mm. The solid lines in figure \ref{fig:4}(b) show the mode frequencies obtained via COMSOL simulations for different volumes of helium accumulated in the device. The color of each line indicates the relative amount of energy stored in the sheath to the energy stored in the funnel. The modes with a large fraction of energy stored in the funnel show dependence on the helium level, decreasing in frequency as the helium level increases

\section{Electrostrictive coupling}
\label{ss:Optomechanical coupling}
In this section the electrostrictive coupling between the optical cavity and the acoustic modes of the liquid helium is described. The coupling arises because the effective cavity length depends on the density of helium that the optical mode overlaps with. Overlap with regions of higher helium density (higher index of refraction) increases the effective length and overlap with the regions of lower helium density (lower index of refraction) decreases the effective length. The expression for the electrostrictive coupling is derived as follows.

Changes in pressure in liquid helium can change the index of refraction locally as seen from the Clausius-Mossotti relation \cite{Abraham1970}:
\begin{equation}
\frac{n^2-1}{n^2+2}=\frac{4\pi\rho\alpha_\mathrm{M}}{3M}
\end{equation}
Here, $n$ is the refractive index, $\rho$ is the density, $\alpha_\mathrm{M}$ is the molar polarizability and $M$ is the molar mass. For $n=1.028$, the left side of the equation is approximately $2(n-1)/3$, and hence $\rho\propto(n-1)$; therefore
\begin{equation}
\frac{\delta \rho}{\rho}=\frac{\delta n}{n-1}
\end{equation}
Given the spacial profile of the relative change in the refractive index $\delta n(\vec{r})$, we can find the change in cavity frequency as:
\begin{equation}
g_0=\omega_c\frac{\int_V I(\vec{r})\delta n(\vec{r})d^3\vec{r}}{\int_V I(\vec{r})n d^3\vec{r}}=\omega_\mathrm{c} (n-1)\frac{\int_V I(\vec{r})\frac{\delta \rho(\vec{r})}{\rho} d^3\vec{r}}{\int_V I(\vec{r})n d^3\vec{r}}
\end{equation}
Here $I(\vec{r})$ is the optical intensity profile and $\omega_\mathrm{c}$ is the optical cavity resonant frequency. The relative change in density for a liquid is equivalent to strain and can be written as a constant $\epsilon_1$ times the dimensionless mode profile $p(\vec{r})$:
\begin{equation}
\frac{\delta \rho(\vec{r})}{\rho}\equiv\epsilon(\vec{r})\equiv\epsilon_1p(\vec{r})
\end{equation}
This leads to the following expression for $g_0$:
\begin{equation}
g_0=\omega_\mathrm{c} (n-1)\epsilon_1\frac{\int_V I(\vec{r})p(\vec{r}) d^3\vec{r}}{\int_V I(\vec{r})nd^3\vec{r}}
\label{eq:g0}
\end{equation}
In the equation \ref{eq:g0} the values for all quantities except for $\epsilon_1$ are known. To find the value of $\epsilon_1$, we write the energy stored in the fluctuations of the mode in terms of the elastic potential energy:
\begin{equation}
E_1=\int_V \frac{1}{2}K\epsilon(\vec{r})^2d^3\vec{r}=\frac{1}{2}v^2\rho\epsilon_1^2\int_Vp(\vec{r})^2d^3\vec{r}
\label{eq:E1pot}
\end{equation}
Here, $K=v^2\rho$ is the bulk modulus, $v$ is the speed of sound. 

To calculate the single photon coupling $g_0$, we equate $E_1$ to the energy stored in a zero point fluctuation:
\begin{equation}
E_0=\frac{\hbar\omega_\mathrm{m}}{4}
\label{eq:E1zero}
\end{equation}
Combining equations \ref{eq:E1pot} and \ref{eq:E1zero}, we solve for the normalization constant $\epsilon_1$: 
\begin{equation}
\epsilon_1=\sqrt{\frac{\hbar\omega_\mathrm{m}}{2 v^2 \rho \int_Vp(\vec{r})^2 d^3\vec{r}}}
\label{eq:epsilon}
\end{equation}
Using equation \ref{eq:epsilon}, the electrostrictive single phonon coupling can be calculated from equation \ref{eq:g0}, given the optical and acoustic mode profiles.

The calculated values of electrostrictive coupling to the radial modes with $m=0$ (red circles) and $m=1$ (blue triangles) vs. the frequencies of the modes obtained using equation \ref{eq:omegamn} are shown in Figure \ref{fig:3}(a). The faint solid lines show the electrostrictive coupling in the case of the optical mode being aligned perfectly with the axis of the cylinder (optical intensity described by equation \ref{eq:Igood}). The bright solid lines show the coupling for the case of optical mode being offset from the axis of the cylinder by $3$ $\mu$m (optical intensity described by equation \ref{eq:Ioffset}). As can be seen from Figure \ref{fig:3}(a), the misalignment results in coupling to $m=1$ mode.

\section{Measurement setup and fits to the data}
\label{ss:Measurement setup}
\begin{figure}[h!] 
\centering
\includegraphics[width=1\textwidth]{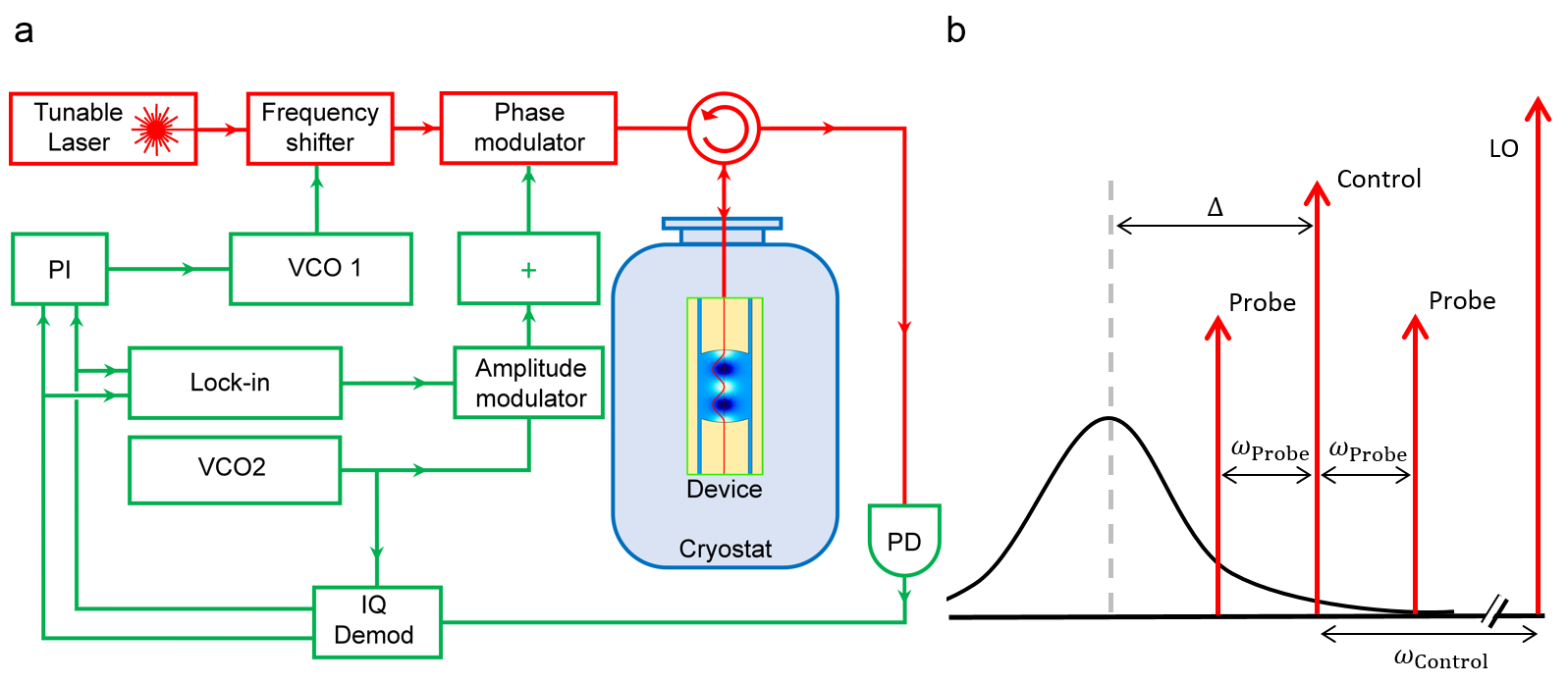} 
\caption{\textbf{The measurement setup}.
\textbf{a)} Schematic of the measurement setup. The optical components are shown in red, electronic components are shown in green. For detailed description see section \ref{ss:Measurement setup}.
\textbf{b)} Illustration of the laser beams incident on the cavity. The LO, control, and both probe beams are shown (red), along with the cavity resonance (black). }
\label{fig:5}
 \end{figure}

\begin{figure}[h!] 
\centering
\includegraphics[width=0.9\textwidth]{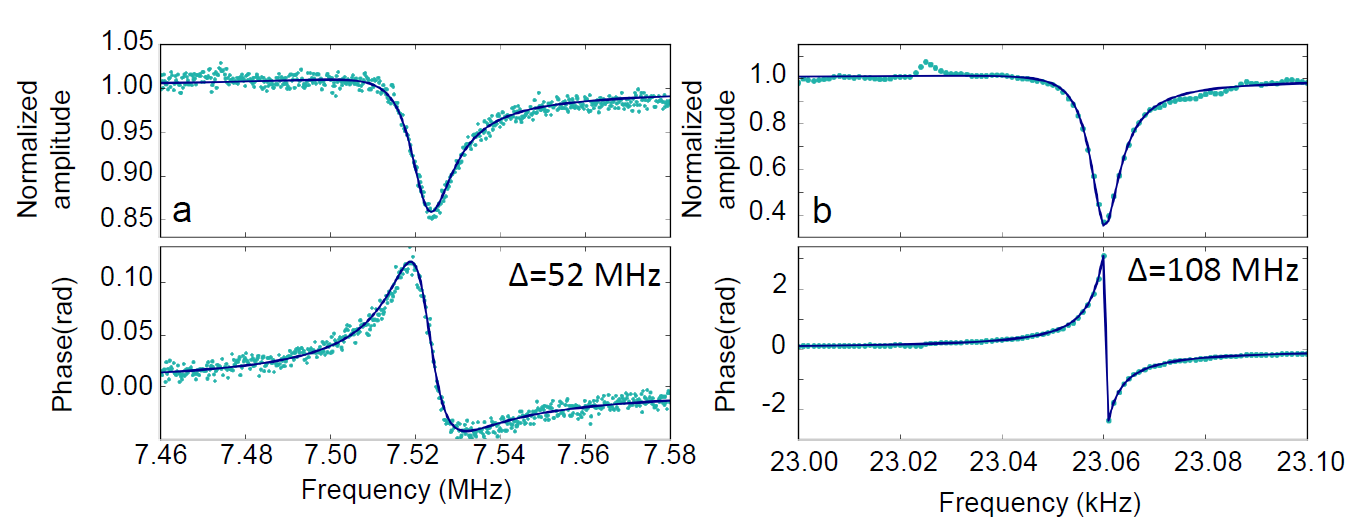} 
\caption{\textbf{OMIT measurements}. The teal points are the normalized amplitude ($a$) and phase ($\psi$) of the OMIT signal as a function of the intracavity beat note frequency $|\omega_{\mathrm{control}}- \omega_{\mathrm{probe}}|/2\pi$. The data is taken by sweeping the frequency difference between the control and probe beams. Shown are the sweeps of the probe beam, which is blue-detuned from the control beam. The data is normalized so that far from the resonance $a = 1$ and $\psi = 0$. The solid blue line is fit to a complex Lorentzian. The fit parameters are $A_+$ and $\Psi_+$ (the overall amplitude and phase of the OMIT effect), $\omega_m$ (the acoustic frequency), and $\gamma_m$ (the acoustic damping rate). \textbf{a)} The sweep over one of the radial modes. The intracavity beat note frequency is varied between 7.46 and 7.58 MHz. The control beam is detuned from the cavity resonance by 52 MHz. \textbf{b)} The sweep over one of the ferrule modes. The intracavity beat note frequency is varied between 23.0 and 23.1 kHz. The control beam is detuned from the cavity resonance by 108 MHz. }
\label{fig:6}
 \end{figure}

The measurement setup is shown in Figure \ref{fig:5}(a). The light leaves a tunable laser and passes through a frequency shifter. The frequency shifter is used to lock the laser frequency to the experimental cavity, as described below. The light then passes through a phase modulator. The phase modulator is used to add three pairs of sidebands: a control beam generated by a Voltage Controlled Oscillator (VCO2) at $\omega_{\mathrm{control}}=2\pi \times 926$ MHz and two probe beams, which are AM sidebands of the control beam and are generated by a microwave amplitude modulator driven at a frequency $\omega_{\mathrm{probe}}$ by a lock-in. 
The carrier beam serves as a local oscillator (LO). The light is delivered to and returned from the cryostat via an optical circulator. Returning light lands on a photodiode. The photocurrent has beatnotes at $\omega_{\mathrm{control}}$ as well as at $\omega_{\mathrm{control}}\pm\omega_{\mathrm{probe}}$. The photocurrent is sent into an IQ demodulator, where it is demodulated at $\omega_{\mathrm{control}}$. Both quadratures on the output of the IQ demodulator have a DC component which carries information about the offset of the control beam from the cavity, and a component at frequency $\omega_{\mathrm{probe}}$, which carries information about the upper and lower sidebands generated by the motion of the acoustic mode.
The DC component is then used to generate a feedback signal to control the frequency shifter via VCO1. Additionally, both quadratures are sent into the lock-in and demodulated at $\omega_{\mathrm{probe}}$ to gain information about the acoustic response at that frequency. For the OMIT measurements the frequency $\omega_{\mathrm{probe}}$ is swept through the resonant frequency of the acoustic mode of interest. Examples of these measurements are shown in Figures \ref{fig:6} (a) and (b).

The data obtained in the manner described above are fit to extract the linewidth and frequency of the acoustic modes, as well the amplitude and phase of the OMIT/OMIA response. All those quantities are fit simultaneously with the OMIT/OMIA theory \cite{Weis2010, Kashkanova2016} to extract the electrostrictive and photothermal coupling. Since the amplitude modulation scheme is employed, the theory described in \cite{Kashkanova2016} needs to be modified to include the second probe beam. This modification is described below.

\subsection{OMIT/OMIA theory with two probe beams}
We start the derivation with the expression for the optical and acoustic amplitudes $\delta \hat{a}[\omega]$ and $\delta \hat{b}[\omega]$ derived in \cite{Kashkanova2016}:
\begin{equation}
\delta{\hat{a}}[\omega] =-i\chi_{\mathrm{cav}}[\omega]\left(g (\delta \hat{b}[\omega]+\delta \hat{b}^\dagger[\omega]) + \sqrt{\kappa_{\mathrm{in}}}\delta s_{\mathrm{in}}[\omega]\right)
\label{eq:a}
\end{equation}

\begin{equation}
\delta \hat{b}[\omega] = \frac{G\sqrt{\kappa_{\mathrm{in}}}(\chi^*_{\mathrm{cav}}[-\omega] \delta s^*_{\mathrm{in}}[\omega]g-\chi_{\mathrm{cav}}[\omega]\delta s_{\mathrm{in}}[\omega]g^*)}{-i(\omega-\omega_\mathrm{m}) + \frac{\gamma_\mathrm{m}}{2}+i\Sigma[\omega]}
\label{eq:b}
\end{equation}
Here $\kappa_{\mathrm{in}}$ is the input coupling, $\delta s[\omega]$ is the amplitude of a probe beam at frequency $\omega$ away from the control beam, $g$ is the multiphoton electrostrictive coupling defined as:
\begin{equation}
g=g_0\frac{-i\sqrt\kappa_{\mathrm{in}}s_{\mathrm{in}}}{-i\bar{\Delta}+\frac{\kappa}{2}}
\label{eq:g}
\end{equation}
 $\chi_{\mathrm{cav}}[\omega]$ is the cavity susceptibility at frequency $\omega$ defined as:
\begin{equation}
\chi_{\mathrm{cav}}[\omega]=\frac{1}{-i\omega- i\bar{\Delta} +\frac{\kappa}{2}}
\label{eq:chi}
\end{equation} 
$\bar{\Delta}$ is the effective detuning of control beam from the cavity and $s_{\mathrm{in}}$ is the amplitude of the control beam. $\Sigma[\omega]$ is optomechanical self-energy defined as:
\begin{equation}
i\Sigma[\omega]=G|g|^2 ( \chi_{\mathrm{cav}}[\omega] -\chi^*_{\mathrm{cav}}[-\omega])
\end{equation}
The $G$ is defined in \cite{Kashkanova2016}:
\begin{equation}
 G=1+\frac{g_{\mathrm{1}}}{g_{\mathrm{0}}}\frac{1}{1-i\omega/\kappa_{\mathrm{Th}}}
 \end{equation}
 
The second term in the definition of $G$ is due to the photothermal coupling. We believe that photothermal coupling arises from the thermal expansion/contraction of the mirrors, due to heating by the laser. The expansion/contraction moves the glass/LHe boundary therefore driving the acoustic mode. The photothermal force has coupling strength $g_1$; $\kappa_{\mathrm{Th}}$ is photothermal bandwidth, and is inversely proportional to the relaxation time for the thermal expansion/contraction of the mirrors.

The optical spring and damping are then defined as: 
\begin{equation}
\Delta \omega_{\mathrm{m}(\mathrm{opt})}=\mathrm{Re}[\Sigma[\omega]]
\label{eq:spring}
\end{equation}
\begin{equation}
\gamma_{\mathrm{m}(\mathrm{opt})}=-2\mathrm{Im}[\Sigma[\omega]]
\label{eq:damping}
\end{equation}

Since an amplitude modulation scheme is employed, the laser has two sidebands: at positive and negative $\Omega$. The expression for $\delta s_{\mathrm{in}}(t)$ is:
\begin{equation}
\delta s_{\mathrm{in}}(t)=s_\mathrm{p}(e^{-i \Omega t} + e^{i \Omega t})
\end{equation}
Taking the Fourier transform and assuming $s_p$ is real:
\begin{equation}
\delta s_{\mathrm{in}}[\omega]=\delta s^*_{\mathrm{in}}[\omega]=\sqrt{2 \pi} s_\mathrm{p}(\delta(\omega-\Omega)+\delta(\omega+\Omega))
\end{equation}
Putting this back into the equation \ref{eq:b}:
\begin{equation}
\delta \hat{b}[\omega] = \frac{\sqrt{2 \pi} s_\mathrm{p} G\sqrt{\kappa_{\mathrm{in}}}(\chi^*_{\mathrm{cav}}[-\omega]g-\chi_{\mathrm{cav}}[\omega]g^*) (\delta(\omega-\Omega)+\delta(\omega+\Omega))}{-i(\omega-\omega_\mathrm{m}) + \frac{\gamma_\mathrm{m}}{2}+i\Sigma[\omega]}
\end{equation}
In the time domain:
\begin{equation}
\delta \hat{b}(t)=b_+[\Omega]s_\mathrm{p} e^{-i \Omega t}+b_-[\Omega]s_\mathrm{p}e^{i \Omega t}
\end{equation}
where 
\begin{equation}
b_+[\Omega]=\frac{G\sqrt{\kappa_{\mathrm{in}}}(\chi^*_{\mathrm{cav}}[-\Omega]g-\chi_{\mathrm{cav}}[\Omega]g^*)}{-i(\Omega-\omega_\mathrm{m}) + \frac{\gamma_\mathrm{m}}{2}+i\Sigma[\Omega]}
\end{equation}
and
\begin{equation}
b_-[\Omega]=\frac{G\sqrt{\kappa_{\mathrm{in}}}(\chi^*_{\mathrm{cav}}[\Omega]g-\chi_{\mathrm{cav}}[-\Omega]g^*)}{-i(-\Omega-\omega_\mathrm{m}) + \frac{\gamma_\mathrm{m}}{2}+i\Sigma[-\Omega]}
\end{equation}
The expression for $b_+$ gives us the motion of the acoustic oscillator; $b_-$ oscillates at $-\Omega$, and so is far off resonance and therefore small.
Writing the acoustic mode amplitude in the time domain and neglecting $b_-$ yields:
\begin{equation}
g(\delta \hat{b}(t)+\delta \hat{b}^\dagger(t))+\sqrt{\kappa_{\mathrm{in}}}\delta s_{\mathrm{in}}(t)=\left((g b_+[\Omega] +\sqrt{\kappa_{\mathrm{in}}})e^{-i \Omega t}+\left(g b_+^*[\Omega]+\sqrt{\kappa_{\mathrm{in}}}\right) e^{i \Omega t}\right)s_\mathrm{p}
\label{eq:deltabt}
\end{equation}
Combining equations \ref{eq:a} and \ref{eq:deltabt}, we express the cavity mode amplitude as:
\begin{equation}
\delta \hat{a}(t)=a_+ [\Omega]e^{-i \Omega t}+a_-[\Omega]e^{i \Omega t}
\end{equation}
where 
\begin{equation}
a_+[\Omega] =-i\chi_{\mathrm{cav}}[\Omega]\left(g b_+[\Omega] +\sqrt{\kappa_{\mathrm{in}}}\right)s_\mathrm{p}
\end{equation}
and 
\begin{equation}
a_-[\Omega] =-i\chi_{\mathrm{cav}}[-\Omega]\left(g b_+^*[\Omega]+\sqrt{\kappa_{\mathrm{in}}}\right)s_\mathrm{p}
\end{equation}
The expression for $a_-$ gives the signal measured at the lower probe beam frequency and the expression for $a_+$ gives the signal measured at the upper probe beam frequency. In Figure \ref{fig:6} (a) and (b), we plot one of the sidebands $a_+$ normalized with respect to the background. The normalized signals $a_-'$ and $a_+'$ are given by:
\begin{equation}
a_-'[\Omega] =\frac{a_-[\Omega]}{a_-[\infty]}=\frac{g b_+^* [\Omega]}{ \sqrt{\kappa_{\mathrm{in}}}}+1
\label{eq:am}
\end{equation}
\begin{equation}
a_+'[\Omega] =\frac{a_+[\Omega]}{a_+[\infty]}=\frac{g b_+ [\Omega]}{ \sqrt{\kappa_{\mathrm{in}}}}+1
\label{eq:ap}
\end{equation}
The normalized signals $a_-'$ and $a_+'$ have Lorentzian shape. The values of amplitude ($A_+$) and phase ($\Psi_+$) of the OMIT/OMIA response at the upper probe beam frequency are defined as the magnitude and phase of $a_+'[\omega_\mathrm{m}]-1$, and the values of amplitude ($A_-$) and phase ($\Psi_-$) of the OMIT/OMIA response at the lower probe beam frequency are defined as the magnitude and phase of $a_-'[-\omega_\mathrm{m}]-1$.

\section{Results and discussion}
\label{s:Results and discussion}

In the Figure \ref{fig:3}(b) the thick black line shows te typical measurement of the OMIT/OMIA response for the intracavity beatnote frequency in the range $1-20$ MHz. The sharp features are the radial modes. The dashed lines are the predicted frequencies for the radial indices $m=0$ (red) and $m=1$ (blue), obtained using COMSOL simulations. The frequencies obtained from COMSOL show good agreement with the frequencies calculated using an analytical expression shown in Figure \ref{fig:3}(a) and with the frequencies obtained experimentally. The features associated with the $m=0$ modes are dominant, but there is also clearly coupling to the $m=1$ modes, which confirms that the optical mode is not perfectly aligned with the axis of the cylinder.

Figure \ref{fig:4}(b) and (c) shows a density plot of the OMIT/OMIA response for intracavity beatnote frequencies in the range $20-300$ kHz (the frequency range of the ferrule modes). The vertical axis is the amount of helium accumulated in the device.

The Figure \ref{fig:4}(b) additionally shows the overlaid COMSOL simulation of the frequencies of the modes In order for the data to agree with the COMSOL simulations, we need to take the volume of helium film to be $V_{\mathrm{film}}=0.22$ cm$^3$, which is within the predicted bounds. The rate at which helium is accumulating in the device is found to be $\dot{V}=6.5\times 10^{-4}$ cm$^3$/min, which is larger than the original prediction of $4\times 10^{-4}$ cm$^3$/min, obtained using the equation \ref{eq:dotv}. The discrepancy can be attributed to the roughness of the inner surface of the capillary, which increases its surface area. 

Additionally, the optical spring and optical damping for both families of modes was observed. The linewidth, frequency, relative amplitude and phase of the OMIT/OMIA feature were measured for the $7.5$ MHz radial mode and the $23$ kHz ferrule mode. 

\begin{figure}[h!] 
\centering
\includegraphics[width=0.9\textwidth]{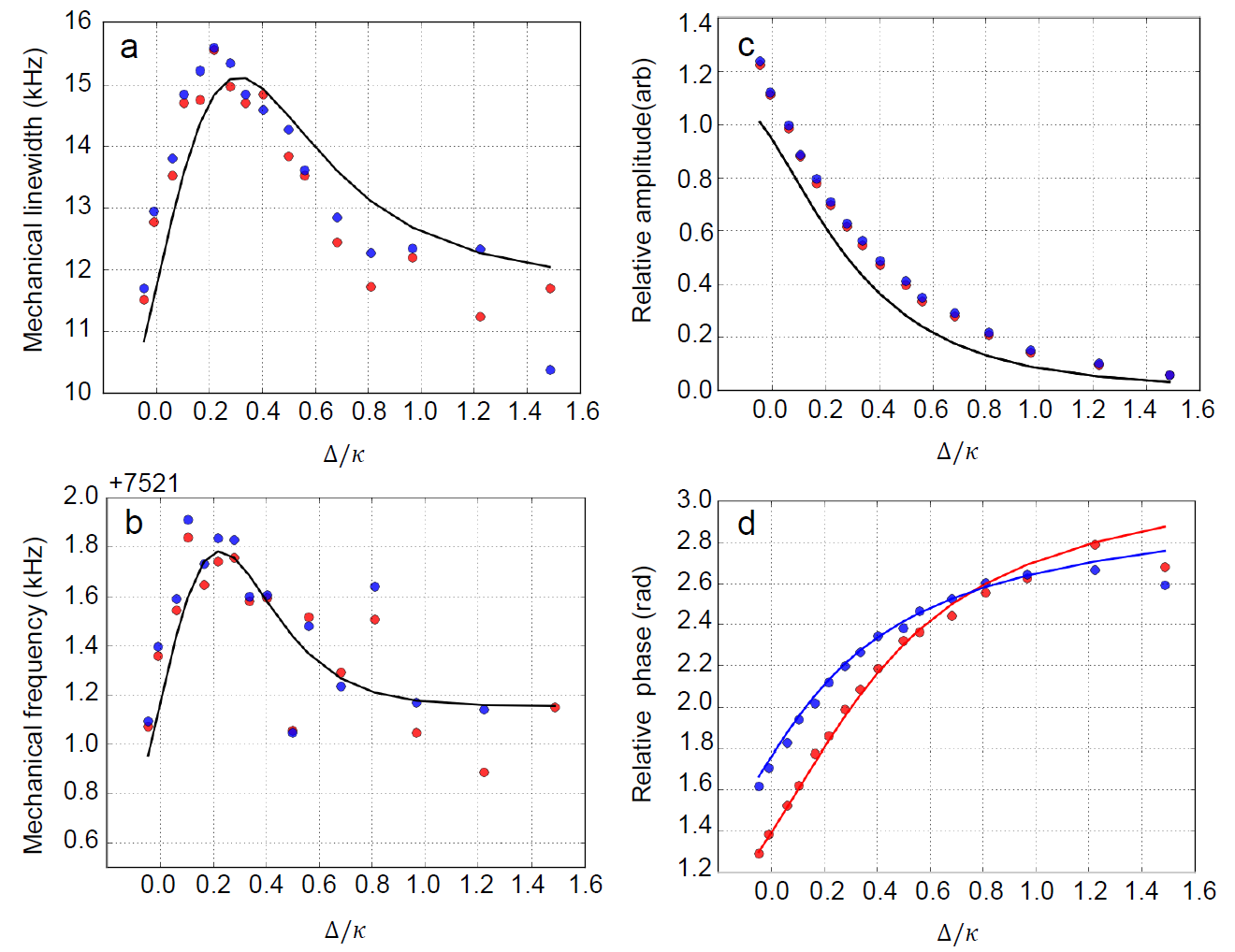} 
\caption{\textbf{Optomechanical effects for the 7.5 MHz radial mode}. Plotted with blue (red) points are the values of the fit parameters for the OMIT/OMIA response at upper (lower) probe beam frequencies extracted from data sweeps such as shown in Figure \ref{fig:6} (a) for various detunings of the control beam. The horizontal axis is the detuning of the control beam from the cavity in units of $\kappa$. The solid lines are the fits discussed in the text, using the measured values of $\kappa$, $\kappa_{\mathrm{in}}$, $s_\mathrm{in}$ and $\bar\Delta$. The fits to the acoustic linewidth, acoustic frequency, amplitude and phase of the OMIT/OMIT response relative to the background are all done simultaneously, assuming the theoretical value of electrostrictive coupling $g_0=2 \pi\times380$ Hz. The values of $\kappa_{\mathrm{Th}}=2\pi\times 1.1 \pm 0.2$ MHz and the ratio of photothermal to electrostrictive coupling strengths $-g_1/g_0=140\pm20$, as well as the the intrinsic linewidth $\gamma_{\mathrm{m}}=2 \pi\times11.7$ kHz and the intrinsic frequency $\omega_{\mathrm{m}}=2 \pi\times7.52$ MHz are the fit parameters extracted from the fit. 
\textbf{a)} The acoustic linewidth. The solid line is the fit to: $(\gamma_{\mathrm{m}}+\gamma_{\mathrm{m (opt)}})/2\pi$ .
\textbf{b)} The acoustic frequency. The solid line is the fit to: $(\omega_{\mathrm{m}}+\Delta \omega_{\mathrm{m (opt)}})/2\pi$.
\textbf{c)} The amplitude of the OMIT/OMIA response relative to the background. The solid line is the fit to $A_+$ and $A_-$ which are the same in the case of amplitude modulation. 
\textbf{d)} The phase of the OMIT/OMIA response relative to the background. The solid blue (red) line is the fit to $\Psi_+$ ($\Psi_-$). } 
\label{fig:7}
 \end{figure}

The data for the radial mode is fit with the OMIT/OMIA theory using the theoretical value of electrostrictive coupling $g_0=2\pi\times380$ Hz, calculated using equation \ref{eq:g0}, assuming the acoustic mode profile $p(r,\theta)=J_0(\alpha_{05} r/R)$. Figure \ref{fig:7} shows plots of linewidth (a), frequency (b), the amplitude (c) and phase (d) of the OMIT feature relative to the background as a function of the control beam detuning for 7.5 MHz mode. It also shows the fits to the theory. We extract the intrinsic linewidth of the mode to be 11.7 kHz, corresponding to a quality factor $Q=642$.
Using the theoretically calculated value for electrostrictive coupling, we find $\kappa_{\mathrm{Th}}=2\pi\times1.1 \pm 0.2$ MHz and the ratio of photothermal to electrostrictive coupling strengths $-g_1/g_0=140\pm20$. 

\begin{figure}[h!] 
\centering
\includegraphics[width=0.9\textwidth]{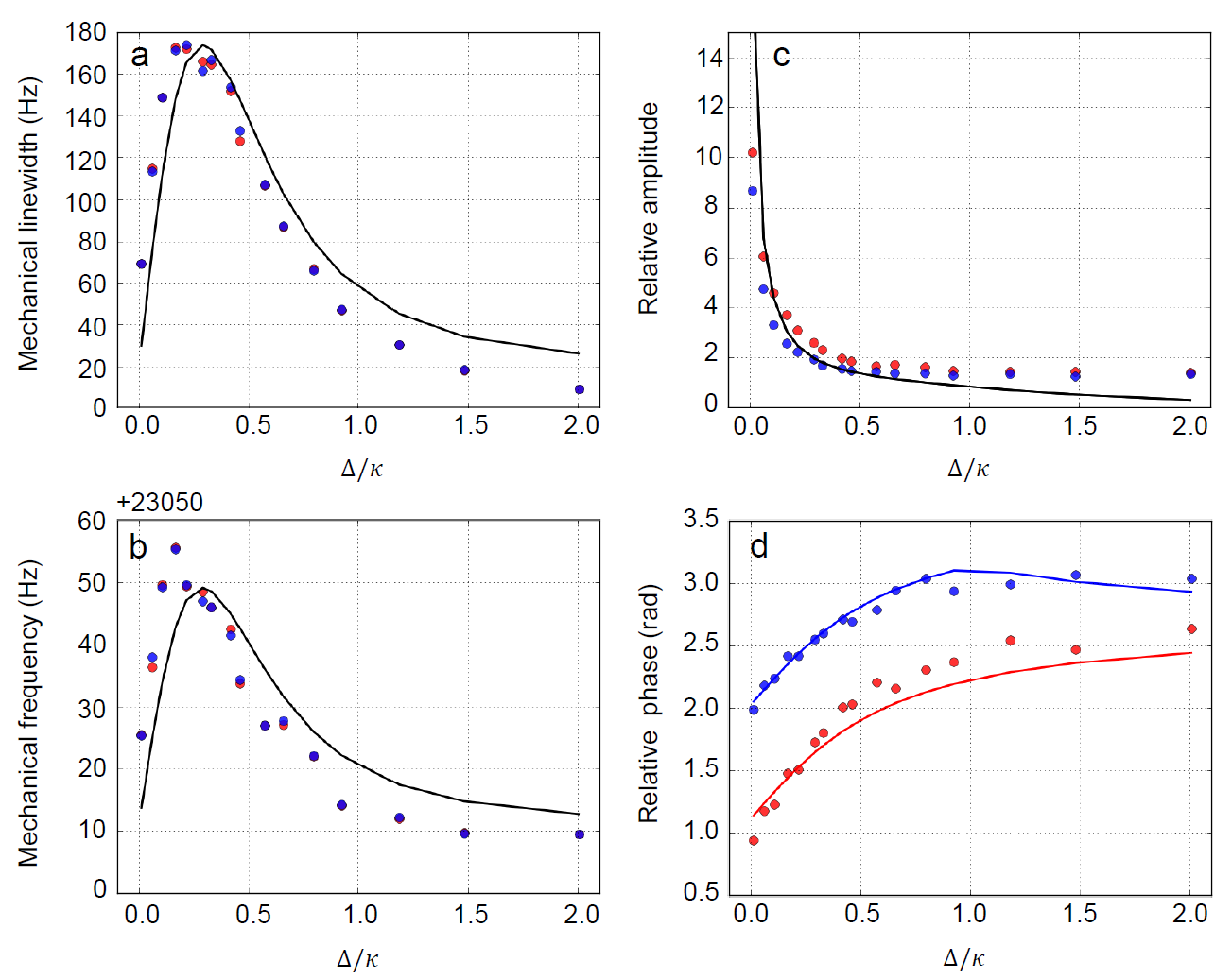} 
\caption{\textbf{Optomechanical effects for the 23 kHz ferrule mode}. Plotted with blue (red) points are the values of the fit parameters for the OMIT/OMIA response at upper (lower) probe beam frequencies extracted from data sweeps such as shown in Figure \ref{fig:6} (b) for various detunings of the control beam. The detuning of the control beam is plotted on the $x$-axis in units of $\kappa$. The solid lines are the theory fits, using the measured values of $\kappa$, $\kappa_{\mathrm{in}}$, $s_\mathrm{in}$ and $\bar\Delta$. The fits to the acoustic linewidth, acoustic frequency, amplitude and phase of the OMIT/OMIA response relative to the background are all done simultaneously. The intrinsic linewidth $\gamma_{\mathrm{m}}=2 \pi\times19$ Hz and the intrinsic frequency $\omega_{\mathrm{m}}=2 \pi\times23.061$ kHz are the fit parameters extracted from the fit. Additionally the value of $\kappa_{\mathrm{Th}}$ is less than $2\pi\times30$ kHz and the electrostrictive coupling rate $g_0$ is between $2 \pi\times230$ Hz and $2 \pi\times600$ Hz as described in the text
\textbf{a)} The acoustic linewidth. The solid line is the fit to: $(\gamma_{\mathrm{m}}+\gamma_{\mathrm{m (opt)}})/2\pi $. 
\textbf{b)} The acoustic frequency. The solid line is the fit to: $(\omega_{\mathrm{m}}+\Delta \omega_{\mathrm{m (opt)}})/2\pi$. 
\textbf{c)} The amplitude of the OMIT/OMIA response relative to the background. The solid line is the fit to $A_+$ and $A_-$ which are the same in the case of amplitude modulation. 
\textbf{d)} The phase of the OMIT/OMIA response relative to the background. The solid blue (red) line is the fit to $\Psi_+$ ($\Psi_-$). } 
\label{fig:8}
 \end{figure}

Figure \ref{fig:8} shows the linewidth (a), frequency (b), amplitude (c) and phase (d) of the OMIT feature relative to the background as a function of the control beam detuning for 23 kHz ferrule mode. The intrinsic linewidth of this mode is 19 Hz, corresponding to a quality factor $Q=1,200$. 

The value of electrostrictive coupling to the ferrule modes is not known $a$ $priori$, so the fits are underconstrained. However large values of $\kappa_{\mathrm{Th}}$ result in almost perfect cancellation of $g_1$ and $g_0$. For example, for $\kappa_{\mathrm{Th}}>2\pi\times30 $ kHz, the ratio $|g_1/g_0+1|<0.05$. Since there is no physical mechanism that would cause the coupling rates to almost cancel, we limit $\kappa_{\mathrm{Th}}<2\pi\times30 $ kHz. This allows us to put bounds on $g_0$ to be between $2\pi\times230$ Hz and $2\pi\times 600$ Hz.


\section{Conclusion}
In the system described, two families of acoustic modes (radial modes and ferrule modes) were observed in addition to the high frequency Brillouin mode described in \cite{Kashkanova2016}. These families are associated with the geometry of the device rather than the cavity. The OMIT/OMIA technique was used to observe the modes. The frequencies of the modes and their dependence on the volume of helium added to the cell were understood. 

Additionally, the modes' optical spring and optical damping were measured. For the radial modes, the electrostrictive coupling rate was calculated. Using this rate, the limits were placed on the photothermal coupling. For the ferrule modes, upper and lower bounds on the electrostrictive coupling rate were found.

\bibliographystyle{plain}
\bibliography{librarynew}
\newpage
\section{Acknowledgements}
We are grateful to Vincent Bernardo, Joe Chadwick, John Cummings, Andreas Fragner, Katherine Lawrence, Donghun Lee, Daniel McKinsey, Peter Rakich, Robert Schoelkopf, Hong Tang, Jedidiah Thompson, and Zuyu Zhao for their assistance. We acknowledge financial support from W. M. Keck Foundation Grant No. DT121914, AFOSR Grants FA9550-09-1-0484 and FA9550-15-1-0270, DARPA Grant W911NF-14-1-0354, ARO Grant W911NF-13-1-0104, and NSF Grant 1205861. This work has been supported by the DARPA/MTO ORCHID program through a grant from AFOSR. This project was made possible through the support of a grant from the John Templeton Foundation. The opinions expressed in this publication are those of the authors and do not necessarily reflect the views of the John Templeton Foundation. This material is based upon work supported by the National Science Foundation Graduate Research Fellowship under Grant No. DGE-1122492. L.H., K.O. and J.R. acknowledge funding from the EU Information and Communication Technologies program (QIBEC project, GA 284584), ERC (EQUEMI project, GA 671133), and IFRAF.
\newpage
\section{Figures}

\end{document}